# Dipole-dipole interaction in random electromagnetic fields

**Sergey Sukhov, Kyle M. Douglass, and Aristide Dogariu***

*CREOL, The College of Optics and Photonics, University of Central Florida,*
*4000 Central Florida Blvd., Orlando, FL,32816, USA*
*\*Corresponding author: adogariu@creol.ucf.edu*

We demonstrate that a non-vanishing interaction force exists between pairs of induced dipoles in random, statistically stationary electromagnetic field. This new type of optical binding force leads to long-range interaction between dipolar particles even when placed in spatially incoherent fields. We also discuss several unique features of dipole-dipole interaction in spatially incoherent Gaussian fields.

It is well known that electromagnetic fields can induce mechanical effects on matter. When an oscillating field **E** impinges on a polarizable object with polarizability $\alpha$, it generates a time-averaged force [1]

$$\overline{\mathbf{F}}(\mathbf{r}) = \tfrac{1}{2}\mathrm{Re}[\alpha^*(\mathbf{r})\mathbf{E}^*(\mathbf{r}) \cdot (\nabla)\mathbf{E}(\mathbf{r})]. \quad (1)$$

This description, valid for harmonic electromagnetic fields, is successfully used to explain numerous experiments involving particles illuminated by coherent laser beams [2].

In many instances however the radiation is far from being fully coherent. A fluctuating electromagnetic field is described by a statistically stationary ensemble of realizations $\mathbf{E}^\gamma = \mathbf{e}^\gamma(\mathbf{r}) U^\gamma(\mathbf{r})$ where $\mathbf{e}^\gamma(\mathbf{r}) = \mathbf{p}^\gamma(\mathbf{r}) + i\mathbf{q}^\gamma(\mathbf{r})$ is the polarization of the realization $\gamma$ of the field and $U$ its corresponding amplitude [3]. Moreover, the electromagnetic fields are in general three-dimensional and, when monochromatic, they can be regarded as an ensemble of plane waves defined by their wavevectors $\mathbf{k}^\gamma$ and their corresponding polarizations $\mathbf{e}^\gamma$ [4]. In these conditions, the force acting on the polarizable object is obtained by averaging the expression in Eq. (1) over the entire ensemble of plane waves representing the fluctuating field. Of course, due to the three-dimensional symmetry, the force exerted on a small object by a three-dimensional, randomly isotropic field averages to zero.

We note that, recently, the effect of spatial and temporal coherence has been discussed in the context of optical forces on microscopic particles [5,6] where the contribution of conservative and nonconservative forces was also estimated. The field-mediated force between two objects in random fields is more subtle. Take, for example, optical binding (OB) [7]. In OB, particles interact via an oscillatory, long-range potential mediated by the field. A number of approaches have been developed that harness the OB force for nanoscale manipulation of matter [2,8]. The initial studies involved fields that were fully coherent, both spatially and temporally, but it was later realized that relaxing the constraints of temporal coherence and using broadband illumination results in a rapid decay of the OB force between two particles. This is due to an overlap of binding potentials for different wavelengths [9,10].

Likewise, one might expect that the OB force averages out in spatially incoherent, fluctuating fields since the oscillatory behavior of the pair interaction potential depends on both the wavevector **k** and the polarization **e** of the incident field [2,8,11]. However, we will show that this is not the case. Remarkably, the OB force between two dipoles survives the ensemble average of field realizations as we will show in the following.

Let us examine the interaction between two induced dipoles located at $\mathbf{r}_1$ and $\mathbf{r}_2 = \mathbf{r}_1 + \mathbf{R}$. The exciting field is quasi-monochromatic, isotropic, and spatially incoherent. This situation describes, for instance, the three-dimensional multiple scattering inside a cavity leading to a Gaussian random field. The field in the system of two identical dipoles can be found self-consistently as [12]

$$\begin{cases} \boldsymbol{\mathcal{E}}^\gamma(\mathbf{r}_1) = \mathbf{E}^\gamma(\mathbf{r}_1) + \overline{\overline{G}}\alpha\boldsymbol{\mathcal{E}}^\gamma(\mathbf{r}_2) \\ \boldsymbol{\mathcal{E}}^\gamma(\mathbf{r}_2) = \mathbf{E}^\gamma(\mathbf{r}_2) + \overline{\overline{G}}\alpha\boldsymbol{\mathcal{E}}^\gamma(\mathbf{r}_1), \end{cases} \quad (2)$$

where $\mathbf{E}^\gamma$ is one realization of the external three-dimensional random field and $\overline{\overline{G}}$ is Green's function tensor [12]. From Eq. (2), one obtains that the field acting on one of the dipoles is

$$\boldsymbol{\mathcal{E}}^\gamma(\mathbf{r}_2) = (\overline{\overline{I}} - \overline{\overline{G}}^2\alpha^2)^{-1}\left[\mathbf{E}^\gamma(\mathbf{r}_2) + \overline{\overline{G}}\alpha\mathbf{E}^\gamma(\mathbf{r}_1)\right]. \quad (3)$$

Because the local fields depend on the locations of the dipoles, the force experienced by dipoles should also depend on their mutual position. In addition, the presence of the second dipole breaks the spherical symmetry of the system and introduces an axis of symmetry along the separation vector **R**. Because of this symmetry, the radial component of the interaction force could survive the average, in contrast to the average force acting on a single dipole, which vanishes upon averaging.

To simplify the notation, we consider the dipoles to be located along the x-axis. The nonvanishing x-component of the interaction force,

$$\langle F_x(\mathbf{r})\rangle = \frac{1}{2}\mathrm{Re}\left(\alpha^*\left\langle \boldsymbol{\mathcal{E}}^{*\gamma}(\mathbf{r})\frac{\partial \boldsymbol{\mathcal{E}}^{\gamma}(\mathbf{r})}{\partial x}\right\rangle_\gamma\right), \quad (4)$$

is calculated by ensemble averaging over all field realizations $\gamma$. After substituting the field given in Eq. (3), one finds that the expression for the force contains linear combinations of different second-order correlations of the excitation field: $\langle |E_u(\mathbf{r})|^2\rangle_\gamma$, $\langle E_u^*(\mathbf{r}_1)E_u(\mathbf{r}_2)\rangle_\gamma$, and $\langle E_u(\mathbf{r}_1)\partial E_u^*(\mathbf{r}_2)/\partial x\rangle_\gamma$ where $u = x, y, z$. For a randomly uniform and isotropic excitation field we find that

$$\langle |E_u(\mathbf{r}_1)|^2\rangle_\gamma = \langle |E_u(\mathbf{r}_2)|^2\rangle_\gamma = \tfrac{1}{3}\langle |\mathbf{E}|^2\rangle_\gamma = \tfrac{1}{3}\langle I\rangle \quad (5)$$

and also

$$\langle E_u^*(\mathbf{r}_1)E_u(\mathbf{r}_2)\rangle_\gamma = \tfrac{1}{3}\langle I\rangle\mu_u(|\mathbf{r}_1-\mathbf{r}_2|) = \tfrac{1}{3}\langle I\rangle\mu_u(R). \quad (6)$$

For Gaussian random fields, the correlation functions $\mu_u(R)$ can be calculated following the approach in Refs. [13,14]. In particular, the transversal $\mu_y(R) = \mu_z(R) = \mu_\perp(kR)$ and the longitudinal $\mu_x(R) = \mu_\parallel(kR)$ correlation functions are found to be

$$\mu_\perp(\xi) = \frac{3}{2}\left(\frac{\sin\xi}{\xi} - \frac{\sin\xi - \xi\cos\xi}{\xi^3}\right),$$
$$\mu_\parallel(\xi) = 3\frac{\sin\xi - \xi\cos\xi}{\xi^3}. \quad (7)$$

To obtain the final expression for the force, we also need to evaluate the two-point correlation function between the field and its gradient along the x-axis: $\nu_{u(x)}(R) = \langle E_u(\mathbf{r}_1)\partial E_u^*(\mathbf{r}_2)/\partial x\rangle$. Following the same approach as in Ref. [13], this field – gradient-of-field correlation can be rewritten as

$$\nu_{u(x)}(R) = \frac{-i}{\langle E^2\rangle}\left\langle (E_u\cdot E_u^*)\frac{k_x}{k}e^{i\mathbf{k}\mathbf{r}}\right\rangle_\gamma. \quad (8)$$

In the case of three-dimensional Gaussian fields one can evaluate the average in Eq. (8) to find the following expressions for the elements of the correlation tensor:

$$\nu_{x(x)}(R) = -3\frac{(3kR\cos(kR) + (-3 + (kR)^2)\sin(kR))}{(kR)^4}, \quad (10)$$
$$\nu_{y(x)}(R) = -3\frac{kR(-3 + (kR)^2)\cos(kR) + (3 - 2(kR)^2)\sin(kR)}{2(kR)^4}.$$

Because of the symmetry $\nu_{y(x)}(R) = \nu_{z(x)}(R)$. Let us note that the correlation function $\nu_{u(x)}(R)$ is anti-symmetric with respect to the sign change in $R$, i.e. $\nu_{u(x)}(R) = -\nu_{u(x)}(-R)$. This also means that $\nu_{u(x)}(0) = 0$ in agreement with the results of Ref.[15] where it was shown that for Gaussian random fields the amplitude of the field and its gradient in the same location are uncorrelated.

Combining all correlation functions in the expression for the force in Eq. (4), we finally obtain the optical binding force acting between two induced dipoles placed in an incoherent field

$$\langle F_x(R)\rangle = \frac{1}{6}\langle I\rangle\left\{-k\,\mathrm{Re}\left(\frac{G_x^*\alpha^{*2}}{1-(G_x^*\alpha^*)^2}\right)\nu_{x(x)}(R) + \right.$$
$$+ |\alpha|^2\,\mathrm{Re}\left(\frac{\partial G_x}{\partial R}\right)\frac{[\mu_x(R)(1+|G_x\alpha|^2) + 2\mathrm{Re}(G_x\alpha)]}{|1-G_x^2\alpha^2|^2} -$$
$$- 2k\,\mathrm{Re}\left(\frac{G_y^*\alpha^{*2}}{1-(G_y^*\alpha^*)^2}\right)\nu_{y(x)}(R) +$$
$$\left. + |\alpha|^2\,\mathrm{Re}\left(\frac{\partial G_y}{\partial R}\right)\frac{2[\mu_y(R)(1+|G_y\alpha|^2) + 2\mathrm{Re}(G_y\alpha)]}{|1-G_y^2\alpha^2|^2}\right\}. \quad (11)$$

Here $G_x = 2\exp(ikR)(-ikR+1)/R^3$ and $G_y = \exp(ikR)(k^2R^2 + ikR - 1)/R^3$ are the two eigenvalues of the Green's function $\overline{\overline{G}}$ [12]. The expression for the binding force in Eq. (11) represents the main result of this Letter. As can be seen, the force depends on the real and imaginary parts of the polarizability $\alpha$ and, therefore, contains contributions from both the conservative gradient forces and from nonconservative radiation pressure.

For small dielectric and nonabsorbing particles, in the limit of weak interaction, i.e. $|G_y\alpha|\ll 1$, $|G_x\alpha|\ll 1$, $\mathrm{Im}\,\alpha \ll \mathrm{Re}\,\alpha$, the expression for the force is considerably simplified:

$$\langle F_x(R)\rangle \approx \frac{1}{6}\langle I\rangle\alpha^2 k^4 f(kR),$$
$$f(\xi) = \frac{3[\xi(18 - 8\xi^2 + \xi^4)\cos(2\xi) + (-9 + 16\xi^2 - 3\xi^4)\sin(2\xi)]}{\xi^7}.$$
(12)

In this case the force depends only on the real part of polarizability $\alpha$, which means that it accounts only for conservative contributions. Moreover, in the limit of small dipoles separation, $\lim_{\xi\to 0} f(\xi) = -(11/5)\xi^{-2}$, the interaction force is attractive. The inverse square law dependence on the separation distance indicates that the near-field interaction is cancelled out from the particle-particle interaction for small separations. In Ref. [16], it was shown that averaging over different orientations of a dipole dimer excited by spatially coherent filed also leads to suppressing the near-field contributions and produces an interaction potential that decays as $1/R$. A similar long-range interaction potential was also found for small separation between pairs of dipoles interacting in Bose-Einstein condensates while being under the excitation of multiple incoherent beams [17]. Let us note that although the approximate expression in Eq. (12) can be rewritten in terms of potential energy of interaction, the full expression given in Eq. (11) cannot because it includes also includes contributions from nonconservative forces proportional to $\text{Im}\alpha$.

For large separation between dipoles excited by a fully coherent field, the OB force decays as $1/R$ while oscillating [12]. When the dipoles are placed in an incoherent field, it follows from Eq.(12) that the forces decay faster, following a $1/R^2$ dependence. This consequence of the optical interaction is noteworthy: even though there is no predominant orientation of the induced dipole moments, the pairwise interaction is still long range. Another interesting observation is that the OB force oscillates at twice the spatial frequency as compared to the case of transversal optical binding [12]. This can be understood from Eq. (11) where the products of two periodic functions (the eigenvalues of $\overline{\overline{G}}$ and the correlation functions $\mu_u(R)$, $\nu_{u(x)}(R)$) results in exponents of the form $\exp(2ikR)$. This means that, in the case of a field with an arbitrary correlation length $r_c$, the spatial oscillations of the resultant interaction force will be determined by the beating of periodic terms proportional to the sum and difference of the two characteristic spatial frequencies, i.e. $2\pi/\lambda$ and $2\pi/r_c$.

In conclusion, we found a non-vanishing optical interaction force between two induced dipoles in a Gaussian, random electromagnetic field. This required evaluating two-point correlation functions between field and its gradient. We found that the particle-particle interaction force is long-range and decays inversely proportional to the square of separation distance between the dipoles. In addition, the magnitude of the force oscillates with a period determined by the two length scale parameters, the wavelength and the random field correlation length.

There are several consequences of our results. For one, it suggests that particle-particle interactions mediated by the electromagnetic field should affect colloidal dynamics in an optically-controlled random medium [18] and the dynamics of atoms in cavity optomechanics [19]. In addition, the interaction between particles subjected to random electromagnetic fields may also provide an experimental testbed for statistical field theories of protein diffusion on membranes [20].